\documentclass[twoside,10pt,a4paper]{newFNLstyle}
\usepackage{graphics}
\usepackage{cite}

\def\w0{\omega_0}

\begin{document}

\volnumpagesyear{5}{2}{000-000}{2005}
\dates{}{}{}

\title{CYCLIC FLUCTUATIONS, CLIMATIC CHANGES AND ROLE OF NOISE IN
PLANKTONIC FORAMINIFERA IN THE MEDITERRANEAN SEA}

\authorsone{A. Caruso$^\diamond$, M.E. Gargano}
\affiliationone{Dipartimento di Scienze della Terra, Universit\`a
di Pisa \\ Via S. Maria 53, I-50100, Italy,
$\diamond$acaruso@dst.unipi.it}

\authorstwo{D. Valenti$^\natural$, A. Fiasconaro and B. Spagnolo}
\affiliationtwo{Dipartimento di Fisica e Tecnologie Relative and
INFM, Group of Interdisciplinary Physics\footnote {Electronic
address: http://gip.dft.unipa.it}, \\Universit\`a di Palermo,
Viale delle Scienze pad. 18, I-90128, Italy,
$^\natural$valentid@gip.dft.unipa.it}

\maketitle

\markboth{A. Caruso, M.E. Gargano, D. Valenti, A. Fiasconaro and
B. Spagnolo}{Climatic Changes and Fluctuations in Zoo-Plankton}

\pagestyle{myheadings}
% Comment this out to remove the running heads


%**************************************************

\begin{abstract}
The study of Planktonic Foraminifera abundances permits to obtain
climatic curves on the basis of percentage ratio between tropical
and temperate/polar forms. Climatic changes were controlled by
several phenomena as: (i) Milankovitch's cycles, produced by
variations of astronomical parameters such as precession,
obliquity and eccentricity; (ii) continental geodynamic evolution
and orogenic belt; (iii) variations of atmospheric and oceanic
currents; (iv) volcanic eruptions; (v) meteor impacts. But while
astronomical parameters have a quasi-regular periodicity, the
other phenomena can be considered as "noise signal" in natural
systems. The interplay between cyclical astronomical variations,
the "noise signal" and the intrinsic nonlinearity of the ecologic
system produces strong glacial or interglacial period according to
the stochastic resonance phenomenon.
\end{abstract}

\section{Introduction}
Planktonic Foraminifera (PF) are unicellular organisms that
commonly live in the sea surface and intermediate water rarely in
the deepest part of water column. PF are very sensible to the
seasonal temperature variations, and in particular some species
prefer sea surface tropical water, while other species prefer
temperate or polar water \cite{Pujol}. The study of dynamic
population of PF permit to obtain climatic curves through two
methods: i) study of the percentage ratio between tropical and
temperate/polar species; ii) analysis of the oxygen isotope
$\delta^{18}O$ variations by carbonatic test of PF
\cite{Shackleton}. PF are in fact good markers for the
reconstruction of Earth climate \cite{Sprovieri, Caruso}. Moreover
other methods are used to reconstruct the Earth's climate history,
as that based on the analysis of variations of ($\delta^{18}O$),
present in the ice cores from Greenland and Antarctic
\cite{Grootes}. According to Refs. \cite{Hilgen}, climatic
fluctuations were essentially produced by cyclical variations of
sun energy received by Earth, that periodically change its
astronomical parameters (equinox precession, 21 ky ($1 ky = 10^3
years$); obliquity of Earth's axis, 41 ky; eccentricity of the
orbit $100$ ky). These astronomical cyclicities are known in earth
sciences as Milankovitch's cycles, and they can be considered a
quasi-deterministic signal that periodically produced drastic
changes in Earth climate. The astronomical forcing is not the only
reason for glacial/interglacial oscillations. In fact insolation
variations produce changes in the atmospheric temperature, but
 the geographical distribution, and position,
of continent and margin plates are extremely important. The
presence of a barrier, as the continental marginal plate, indeed
can influence the atmospheric and oceanic currents. In addition to
these global events, other "randomic" events occur: (i) tectonic
uplift of belt ridge; (ii) volcanic activity, in particular
explosive eruptions; (iii) meteor impacts, that occasionally
strike the Earth and that caused several catastrophic mass
extinctions \cite{Alvarez}. The Mediterranean Sea, because of its
geographical position, between tropical and temperate area, is a
good laboratory for the study of the climatic variations of Earth
history \cite{Caruso}. High resolution studies carried out on
marine sediments and ice cores from Greenland and Antarctic, have
demonstrated that in the last $400$ kyr, in addition to the
classic Milankovitch cycles, other cyclic variations were present
with a higher frequencies, known as sub-Milankovian cycles. In
particular spectral analysis carried out on these sequences has
permitted to recognize periodicities of $5000$, $2500$, $1600$ and
$200$ years \cite{Alley}. Recently geochemical isotopes of ice
cores from Greenland have suggested that Earth's climate
variations occur according to the model based on stochastic
resonance phenomenon \cite{Alley}.

\section{Experimental data}
In this work we analyze data from: (i) Mediterranean core
sediments (Sites $963$, Leg $160$), taken during Ocean Drilling
Program (ODP), and compared with upper Pleistocene ice cores (GRIP
and GISP2), from Greenland \cite{Grootes}; (ii) landscape section
outcropping in Southern Sicily, and compared with
Pliocene-Pleistocene core sediments come from Atlantic Ocean Sites
$659$ \cite{Haug}. For detailed description of experiments see
Refs.~\cite{Sprovieri,Caruso,Haug}.

\begin{figure}[htbp]
\centering{\resizebox{8cm}{!}{\includegraphics{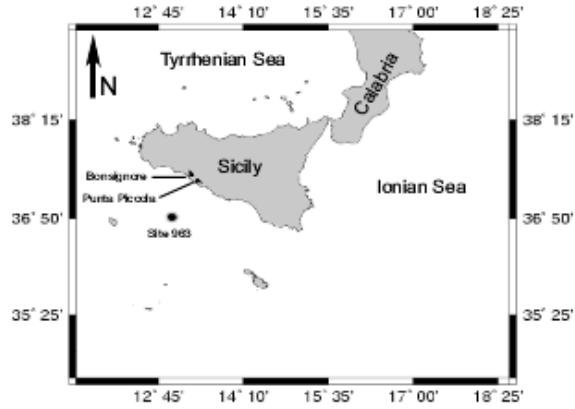}}}
\vskip -0.6cm
 \caption{Location map of studied sections in the
Mediterranean area}\label{Fig.1}
\end{figure}

\begin{figure}[htbp]
\centering{\resizebox{10 cm}{!}{\includegraphics{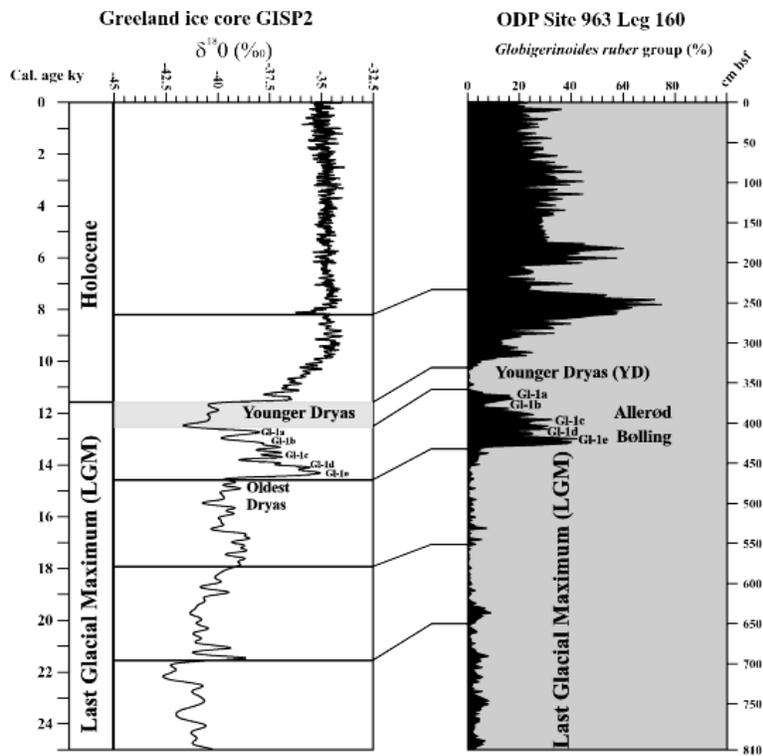}}}
\caption{Correlation between oxygen isotope $\delta^{18}O$ (GISP2)
and percentage fluctuations of \emph{G. ruber} of Site $963$. The
Last Glacial Maximum (LGM) and Younger Dryas (YD) correspond to
minima percentages of \emph{G. ruber}. This tropical species is
always present in the Mediterranean area from today until $\simeq
10$ ky (B.P.), but during glacial phases was absent or very rare.}
\label{Fig.2}
\end{figure}

\begin{figure}[htbp]
\centering{\resizebox{10cm}{!}{\includegraphics{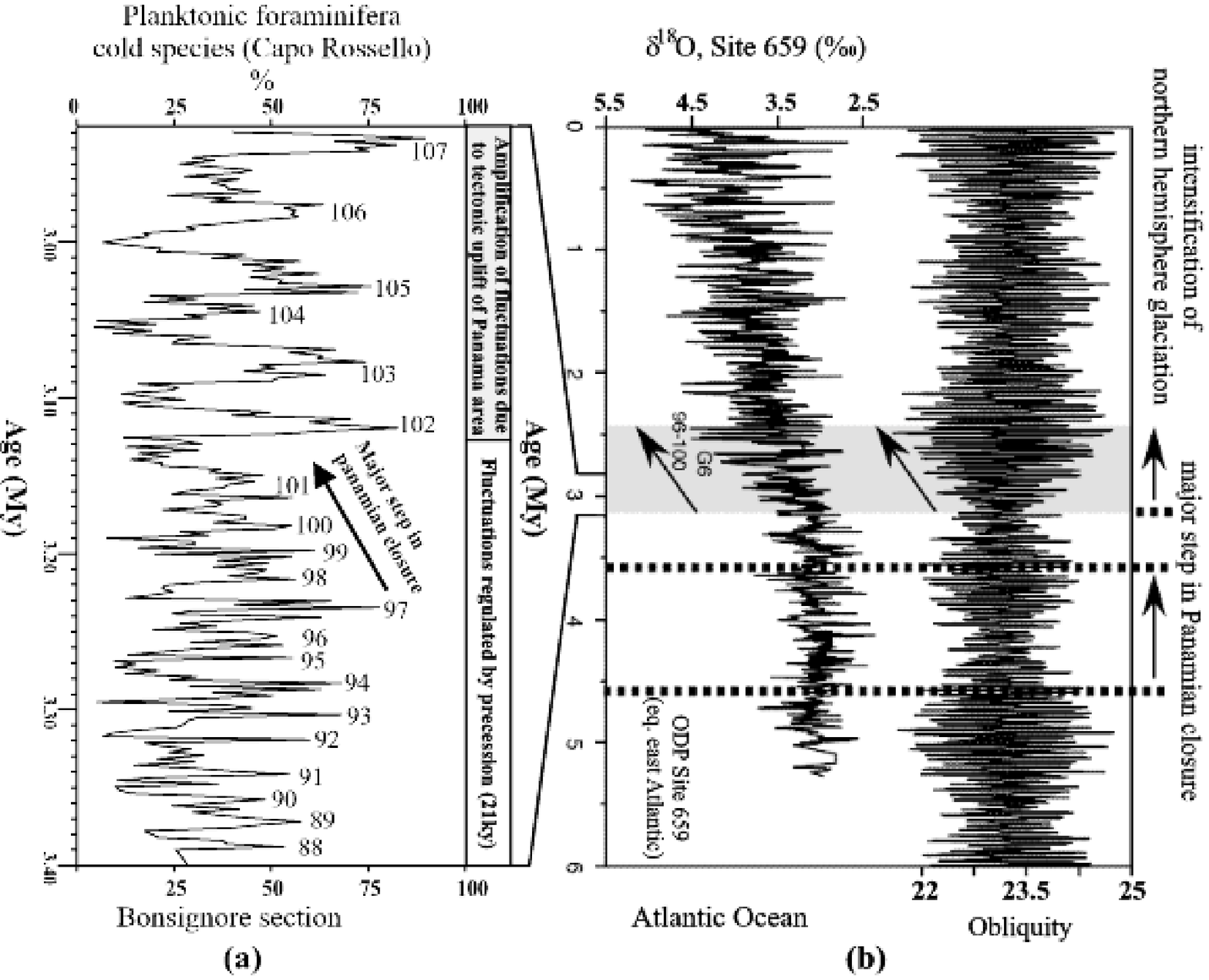}}}
\vskip -0.6cm \caption{(a) Cold species (\emph{G. bulloides})
fluctuations between $3.4$ My and $2.93$ My (Bonsignore section)
and (b) comparison with oxygen isotope ($\delta^{18}O$) of Site
$659$ \cite{Haug} between today and $6$ My. The number of
fluctuations $88-107$ correspond to lithological cycles of Punta
Piccola described by \cite{Hilgen}. The increase of the amplitude
of oscillations of cold species in (a) coincides with increase of
tectonic uplift of Panama area in (b).}
\vskip -0.6cm
\label{Fig.3}
\end{figure}
(i) \textbf{\emph{Mediterranean ODP sites}} - The core from Site
963 (central part of Mediterranean Sea, Sicily Channel),
consisting of grey marls rich of PF,  was sampled each $2$ cm in
order to have a continuous sequence from today to $25$ ky before
present (B.P.). The PF fluctuation abundances \cite{Sprovieri} are
compared with oxygen isotope of Gisp2. There is a good correlation
between the fluctuation percentage ratio of a particular warm
species (\emph{Globigerinoides ruber}) and $\delta^{18}O$
variations. In particular while \emph{G. ruber} is always present
in the interval today-$10$ ky, even if small oscillations in
percentage ratio have been observed, in some particular intervals,
between $14.5$ ky and $25$ ky, \emph{G. ruber} is absent or very
rare (Fig.~2) due to the drastic decrease of global temperature
during Younger Dryas and Last Glacial Maximum periods.

(ii) \textbf{\emph{Southern Sicily}} - The Bonsignore section
(near Ribera) is characterized by a continuous sequence of
marly-limestone and marls of Trubi and Monte Narbone Formations
\cite{Caruso,Hilgen}, rich in PF. This sedimentary sequence is
coeval with Punta Piccola section (Agrigento), where lithological
cycles $88-107$ were described \cite{Hilgen}. Bonsignore section
recover a time interval between $3.4$ My (million years ago) and
$2.93$ My. From $3.4$ to $3.2$ My, lithological and foraminifera
cyclicities are mostly controlled by precession forcing (Fig.~3a),
while after $3.2$ My the obliquity forcing becomes dominant
\cite{Caruso,Hilgen}. This change in climatic Earth system is
connected with a casual event as the uplift of Panama area and,
consequently, with the closure of Panamian Isthmus. Fig.~3a
represents the time interval around the major step in Panama
closure. This event produced a drastic change in oceanic and
atmospheric circulation with the amplification of Gulf Current
gyre and an increase of the formation of North Atlantic Deep Water
(NADW) in consequence \cite{Haug, Caruso}. In Fig.~3 it is
possible to observe that the drastic increase of oscillation
amplitude of cold species, recognized in the Mediterranean Sea,
starts in coincidence with sharp increase of oxygen isotope
(Atlantic Ocean). The Mediterranean Sea has perfectly recorded
this important global climatic cooling that favored the formation
of Polar Ice Sheet \cite{Haug, Caruso}.

\section{The model}

The dynamics of the biological system above described appear
rather complex due to the presence of periodicities which
sometimes disappear. The main peculiarities observed from
experimental time series of PF are: (i) geological events produce
"time windows" characterized by quasi-periodic fluctuations with
almost constant intensity, which can be ascribed to different
\emph{"noise levels"}; (ii) some periodicities appear in one of
this "time window", while are absent in all the other ones; (iii)
the two species \emph{G. ruber} and \emph{Globigerina bulloides}
coexist in a competing dynamical regime. As a first approximation
we try to describe the behavior of our ecosystem by a stochastic
model of two competing species by using the following generalized
Lotka-Volterra (LV) equations \cite{Spagnolo,Valenti}

\vskip -0.6cm
\begin{eqnarray}
\dot{x}=\thinspace x\thinspace(1-x-\beta(t) y)+
x\thinspace\xi_x(t)\\
\dot{y}=\thinspace y\thinspace(1-y-\beta(t)
x)+y\thinspace\xi_y(t),
 \label{LotVol}
\end{eqnarray}
where $\xi_x(t)$ and $\xi_y(t)$ are statistically independent
$\delta$-correlated Gaussian white noises with zero mean. The
multiplicative noise models the interaction between the
environment and the species. The interaction parameters $\beta$ is
characterized by a critical value corresponding to $\beta_c = 1$.
For $\beta < \beta_c$ a coexistence regime of the two species is
established, while for $\beta > \beta_c$ an exclusion regime takes
place, i.e. in a finite time one of the two species extinguishes.
It is then interesting to investigate the time evolution of the
ecosystem for $\beta$ varying around the critical value $\beta_c$
in the presence of fluctuations, due to the significant
interaction with the environment. This behavior can be obtained
assuming $\beta$ subjected to a bistable potential and a periodic
driving according to the following stochastic differential
equation \cite{Valenti}
 \begin{equation}
\frac{d\beta(t)}{dt} = -\frac{dU(\beta)}{d\beta}+\gamma
cos(\omega_0 t) + \xi_{\beta}(t),
\label{beta_eq}
\end{equation}
where $\gamma=10^{-1}$, $\omega_0/(2\pi)=10^{-3}$, and $U(\beta)$
is a generalized bistable potential $U(\beta) =
h(\beta-(1+\rho))^4/\eta^4-2h(\beta-(1+\rho))^2/\eta^2$. The
stable states correspond to the two regimes of the deterministic
LV model. In Eq.(\ref{beta_eq}) $\xi_{\beta}(t)$ is a
$\delta$-correlated Gaussian white noise with zero mean.
\begin{figure}[htbp]
\centering{\resizebox{10cm}{!}{\includegraphics{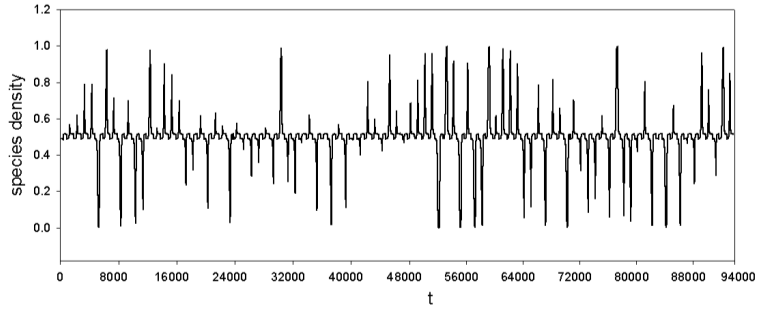}}}
\vskip -0.3cm \caption{Time evolution of both populations at
different levels of the multiplicative noise, namely
$\sigma=10^{-10}$ for $0<t<48000$, and $\sigma=10^{-9}$ for
$t>48000$. The values of the parameters are $\gamma = 10^{-1}$,
$\w0/2\pi = 10^{-3}$, and $\sigma_\beta=1.78 \cdot 10^{-3}$. The
initial values of the two species are $x(0)=y(0)=1$.}
\vskip -0.3cm
\label{time_series}
\end{figure}
To analyze the dynamics of the two species we fix the additive
noise intensity at the value $\sigma_\beta=1.78\cdot 10^{-3}$
\cite{Valenti}. The time series of the two species are obtained
for two different values of the multiplicative noise intensity
$\sigma=10^{-10}, 10^{-9}$ (see Fig.~\ref{time_series}). For
$\sigma=10^{-10}$ the two species coexist and quasi-periodic
oscillations appear with random periodical inversions of
populations (Fig.~\ref{time_series}). An increase of the noise
($\sigma=10^{-9}$) produces an enhancement of the amplitude of
these quasi-periodical oscillation as observed in experimental
data (see Fig.~3). The appearance of some periodicities,
previously "hidden", are due to the stochastic resonance
phenomenon. The periodical signal of small amplitude, that is the
obliquity in Fig.~3, is enhanced by the presence of the noise
~\cite{Valenti,Benzi}. We note finally that the theoretical model,
based on the stochastic resonance phenomenon, predicts a time
behavior of the two species abundances, which cannot be obtained
by using models which are simply periodic or stochastic
\cite{Spagnolo,Valenti,Spagnolo1}.

\section{Conclusions}

 The main peculiarities observed by analyzing our experimental data of
PF in Sicily Channel can be explained within the proposed model of
SR in population dynamics \cite{Valenti}. The nonlinearity of the
natural system together with a periodical forcing and a
"\emph{noise signal}" produces a coherent response of the
ecosystem, by enhancing the effect of the geological causes.

This work has been supported by INTAS Grant 01-450, INFM and MIUR,
and by Grant n. 77502 "Geosites of Sicily" of Sicily Region, A. B.
Cult. ed Ambientali.

\end{document}